\documentclass[%
 reprint,
superscriptaddress,
nobibnotes,
 amsmath,
 amssymb,
 aps,
prb,
floatfix,
]{revtex4-2}
\usepackage{hyperref} 
\hypersetup{ colorlinks=true, linkcolor=blue, filecolor=violet, urlcolor=blue ,citecolor=blue}

\usepackage{graphicx}% Include figure files
\usepackage{dcolumn}% Align table columns on decimal point
\usepackage{bm}% bold math
\usepackage[utf8]{inputenc}% http://ctan.org/pkg/inputenc
\usepackage{amsmath}

\begin{document}
\preprint{APS/123-QED}

\title{Polarimetric analysis of thermal emission from both reciprocal and nonreciprocal materials using fluctuational electrodynamics}% Force line breaks with \\

\author{Chiyu Yang}
 \affiliation{George W. Woodruff School of Mechanical Engineering, Georgia Institute of Technology, Atlanta, GA 30332, USA}
\author{Wenshan Cai}
 \affiliation{School of Electrical and Computer Engineering, Georgia Institute of Technology, Atlanta, GA 30332, USA}
\author{Zhuomin M. Zhang}
\thanks{zhuomin.zhang@me.gatech.edu}%
 \affiliation{George W. Woodruff School of Mechanical Engineering, Georgia Institute of Technology, Atlanta, GA 30332, USA}

\begin{abstract}
Coherent thermal emission for a given polarization has been observed in many metamaterials with micro/nanostructures. A complete description of the thermal emission requires the full characterization of the spectral angular emissivity for all polarization states. Emissivity is typically obtained based on the equivalence between the absorptivity and emissivity according to Kirchhoff's law; however, such relation may be invalid for nonreciprocal media. More general approaches without the constrain of optical reciprocity are necessary when dealing with magneto-optical materials and magnetic Weyl semimetals. Here, a polarimetric analysis of thermal emission is carried out based on fluctuational electrodynamics. The Stokes parameters are obtained using coherency matrix for a multilayered system with anisotropic media, including nonreciprocal materials. The results demonstrate that thermal emission may be circularly or linearly polarized in different directions and frequencies. The findings are consistent with the statements of the modified Kirchhoff's law provided by several groups in recent years, and therefore, justify the appropriateness of both the direct and indirect methods. This study will help the design of desired thermal emitters for energy harvesting and thermal control. 

\end{abstract}

%Use showkeys class option if keyword  %display desired

\maketitle

\section{Introduction}
Traditionally, thermal radiation from solid materials is thought of as incoherent, broadband, and unpolarized; examples are incandescent lamps, hot plates, bricks, and even human bodies \cite{RN1}. Wavelength-selective emitters have been extensively studied in recent years using micro/nanostructured materials \cite{RN2,RN3}. Modification of thermal emission has numerous practical applications such as (i) tuning the emission spectrum helps improve the efficiency of thermophotovoltaic systems \cite{RN4}, (ii) controlling the emission direction benefits radiative cooling and waste heat recovery \cite{RN5}, and (iii) the polarization facilitates biomedical diagnostics and target detecting \cite{RN6}. Directional thermal emission with a specified linear polarization has been observed in many metamaterials supporting electromagnetic resonances, e.g., surface plasmon polaritons, surface phonon polaritons, magnetic polaritons, and surface waves in photonic crystals \cite{RN7, RN8, RN9, RN10}. Generally speaking, the emissivity of metamaterials is spectral, angular, and polarized dependent. Therefore, it is of great importance to fully describe the spectral angular emissivity and polarization states of the emitters. 

Spectral angular emissivity is commonly obtained by its equivalence to spectral angular absorptivity according to Kirchhoff's law. For an opaque object, since the absorptivity is related to the reflectance, the indirect method allows one to obtain the emissivity from the reflectance. However, as demonstrated recently, Kirchhoff's law does not hold for all thermal emitters, e.g., the nonreciprocal emitters made by a magneto-optical material \cite{RN11, RN12} or a Weyl semimetal (WSM) \cite{RN13}. These nonreciprocal materials may allow circularly polarized thermal emission to be realized and holds great promise to further improve the performance of solar energy converters \cite{RN14}. Therefore, a more general approach is needed to characterize the emissivity without restricting to optical reciprocal materials or structures. Several recent studies have provided deeper understandings and generalizations of the traditional Kirchhoff's law \cite{RN11, RN12, RN13, RN14, RN15, RN16}. Zhu and Fan \cite{RN11} demonstrated the nearly complete violation of Kirchhoff's law. Zhang et al. \cite{RN12} discussed Kirchhoff's law from the points of co- and cross-polarization energy balance. Khandekar et al. \cite{RN15} described the relationship between circularly polarized emissivity and absorptivity. Guo et al. \cite{RN16} presented an adjoint Kirchhoff's law that relate the emissivity of an object to the absorptivity of its adjoint system. Most of these works used indirect method to calculate thermal emissivity. Although direct method based on fluctuational electrodynamics has been used to predict the thermal emission \cite{RN11, RN17, RN18}, a comprehensive analysis of polarized emissivity using the direct method to calculate thermal emission, especially for the nonreciprocal case, is imperative. 

In this work, fluctuational electrodynamics is used as a direct approach to calculate thermal emission from multilayer structures where each layer may be isotropic or anisotropic (whether reciprocal or not) as long as it is nonmagnetic. Electromagnetic field is given in the form of coherency matrix by using fluctuation-dissipation theorem (FDT) with the help of dyadic Green's function (DGF). The coherency matrix is written as a function of a wave vector that corresponds to the direction of emission. To describe the polarization states, Stokes parameters are obtained according to the elements in the coherency matrix; and the polarized emissivities are determined consequently. Several multilayered structures are designed and analyzed to predict the spectral, angular, and polarization-dependent emissivity. These examples are intended not only to demonstrate the methodology but also to explore exotic radiative properties enabled by anisotropic multilayered structures.

\section{Theory}
\subsection{Fluctuational electrodynamics}

Fluctuational electrodynamic is a powerful tool for calculating thermal radiation from a body in both the near field and far field. Consider a nonmagnetic medium in a vacuum with a tensorial form of relative permittivity $\tensor{\varepsilon}$. At thermal equilibrium with a temperature $T$, the induced electric field $\bf{E} (\bf{r},\omega)$ in free space is expressed by the induced current density $\bf{J} (\bf{r'},\omega)$ in the medium with the help of the dyadic Green's function $\tensor{G}$ as \cite{RN2}
%%%%%%%%%%%%%%%
\begin{equation} \label{eq1}
\textbf{E} (\textbf{r},\omega) = i\omega\mu_0\int \tensor{G} ({\bf{r}},{\bf{r'}},\omega ) \cdot {\bf{J}}({\bf{r'}},\omega ){d}{\bf{r'}}
\end{equation}
%%%%%%%%%%%%%%%%
where $\bf{r}$ and $\bf{r'}$ denote the locations of the resultant electric field and the source current density, respectively, $\omega$ is the angular frequency, $i$ is the unit imaginary number, and $\mu_0$ is the vacuum permeability. The correlation function of the electric field at location ${\bf{r}}_1$ and ${\bf{r}}_2$ is given by a double integration over the volume containing the current source \cite{RN19}: 
%%%%%%%%%%%%%%%%
\begin{equation} \label{eq2}
\begin{split}
\langle {\bf{E}}({\bf{r}} _1,\omega ) \otimes {\bf{E}}^*({\bf{r}}_{2},\omega ) \rangle =&  \langle \mu_0^2 \omega ^2 \iint {\tensor{G}} ({\bf{r}}_1,{\bf{r'}},\omega )  {\bf{J}}({\bf{r'}},\omega ) \\&\otimes{\bf{J}}^*({\bf{r''}},\omega){\tensor{G}}^\dag  ({\bf{r}}_2,{\bf{r''}},\omega ){d}{\bf{r'}}{d}{\bf{r''}}   \rangle
\end{split}
\end{equation}
%%%%%%%%%%%%%%%%
where $\langle\rangle$ stands for ensemble averaging, ${\bf{a}} \otimes {\bf{b}} = {\bf{a}}{{\bf{b}}^{\rm{T}}}$, $*$ denotes complex conjugate, and $\dagger$ signifies conjugate transpose. The current density correlation function is given by FDT as \cite{RN19, RN20, RN21}
%%%%%%%%%%%%%%%%%%%%%
\begin{equation} \label{eq3}
\begin{split}
\langle {{\bf{J}}({\bf{r'}},\omega ) \otimes {{\bf{J}}^*}({\bf{r''}},\omega )} \rangle  =& \frac{4}{\pi }\omega {\varepsilon _0}\Theta (\omega ,T)\\&\times\frac{{\mathord{\buildrel{\lower3pt\hbox{$\scriptscriptstyle\leftrightarrow$}} 
\over \varepsilon } (\omega ) - {{\mathord{\buildrel{\lower3pt\hbox{$\scriptscriptstyle\leftrightarrow$}} 
\over \varepsilon } }^\dag }(\omega )}}{{2i}}\delta ({\bf{r'}} - {\bf{r''}})
\end{split}
\end{equation}
%%%%%%%%%%%%%%%%%%%%
Here, $\varepsilon _0$ is the vacuum permittivity, $\Theta (\omega ,T) = \hbar \omega /2 + \hbar \omega /(\exp (\hbar \omega /{k_{\rm{B}}}T) - 1) $ is the mean energy of Planck's oscillator, and $\delta$ is a Dirac delta function. A factor 4 is multiplied to the expression since only positive frequencies are considered. Once the DGFs are calculated, the local energy density, optical intensity, and Poynting vector can all be evaluated \cite{RN19}. The current density correlation function provided in Eq. (\ref{eq3}) is known as the second kind of FDT \cite{RN18, RN22}.

While DGFs have been successfully applied in many studies dealing with isotropic media and relatively simple geometric structures, the evaluation of DGFs in Eqs. (\ref{eq1}) and (\ref{eq2}) for a medium with tensorial permittivity is mathematically intensive due to the complex dispersion relations. Only in special cases, such as with a uniaxial medium whose  optic axis is aligned with one of the coordinates, DGFs that directly relate the current source and resultant field can be explicitly obtained \cite{RN23}. An alternative approach based on scattering theory is often adopted to facilitate the computation of DGFs involving anisotropic or more complicated geometries. If the system is at global thermal equilibrium, i.e., both the vacuum and the medium are at the same temperature, an important identity can be applied to the DGFs \cite{RN24,RN25}:
%%%%%%%%%%%%%%%%%%%%%%%%
\begin{equation} \label{eq4}
\begin{split}
\frac{\omega ^2}{c^2} \int \tensor{G}({\bf{r}}_1,{\bf{r'}},\omega ) 
&\frac{\tensor{\varepsilon} (\omega )-\tensor{\varepsilon} ^\dag (\omega )} {2i} 
\tensor{G}^\dag({\bf{r}}_2,{\bf{r'}},\omega ){d}{\bf{r'}}
\\& = \frac{\tensor{G} ({\bf{r}}_1,{\bf{r}}_2,\omega) - {\tensor{G}}^\dag ({\bf{r}}_2,{\bf{r}}_1,\omega )}{{2i}}
\end{split}
\end{equation}	
%%%%%%%%%%%%%%%%%%%%%%
This identity holds true for both reciprocal and nonreciprocal materials. Hence, the electric field correlation function at global equilibrium is reduced to \cite{RN26,RN27} 
%%%%%%%%%%%%%%%%%%%%%%%%%%%
\begin{equation} \label{eq5}
\begin{split}
\langle {{{\bf{E}}_{{\rm{ge}}}}({{\bf{r}}_1},\omega ) \otimes {{\bf{E}}_{{\rm{ge}}}^*}({{\bf{r}}_2},\omega )} \rangle  =&  \frac{4}{\pi }\omega {\mu _0}\Theta (\omega ,T) \\&\times\frac{{\tensor{G}} ({{\bf{r}}_1},{{\bf{r}}_2},\omega ) - {{\tensor{G} }^\dag }({{\bf{r}}_2},{{\bf{r}}_1},\omega )}{{2i}}
\end{split}
\end{equation}	 
%%%%%%%%%%%%%%%%%%%%%%%%%%
where the subscript “ge” stands for global equilibrium. This relation and its similar forms are called the first kind of FDT that requires the system at global thermal equilibrium \cite{RN18,RN22}. Since both positional variables ${\bf{r}}_1$ and ${\bf{r}}_2$ of the DGF in Eq. (\ref{eq5}) are located in free space, it can be derived relatively easily based on the scattering approach due to the simple dispersion relation of vacuum. For a multilayer structure, the DGF can be written in terms of Fresnel's reflection coefficients at the vacuum-object interface. At global thermal equilibrium, the electric field and the field correlation may be decomposed into contributions from the medium and vacuum, since there is no correlation between the two. As a result, field correlation in Eq. (\ref{eq2}) can be obtained by subtracting the vacuum component from the global equilibrium term expressed in Eq. (\ref{eq5}) as follows \cite{RN18,RN28}
%%%%%%%%%%%%%%%%%%%%%%%%%%%
\begin{equation} \label{eq6}
\begin{split}
\langle {{\bf{E}}({{\bf{r}}_1},\omega ) \otimes {{\bf{E}}^*}({{\bf{r}}_2},\omega )} \rangle = &  \langle {{{\bf{E}}_{{\rm{ge}}}}({{\bf{r}}_1},\omega ) \otimes {{\bf{E}}_{{\rm{ge}}}^*}({{\bf{r}}_2},\omega )} \rangle \\ &- \langle {{{\bf{E}}_{{\rm{vac}}}}({{\bf{r}}_1},\omega ) \otimes {{\bf{E}}_{{\rm{vac}}}^*}({{\bf{r}}_2},\omega )} \rangle 
\end{split}
\end{equation}	 
%%%%%%%%%%%%%%%%%%%%%%%%%%
where the subscript “vac” represents vacuum. One can apply the similar approach to calculate near-field radiative heat transfer between many bodies at different temperatures \cite{RN27,RN29}.

\begin{figure}[!ht]
\includegraphics[width=86mm]{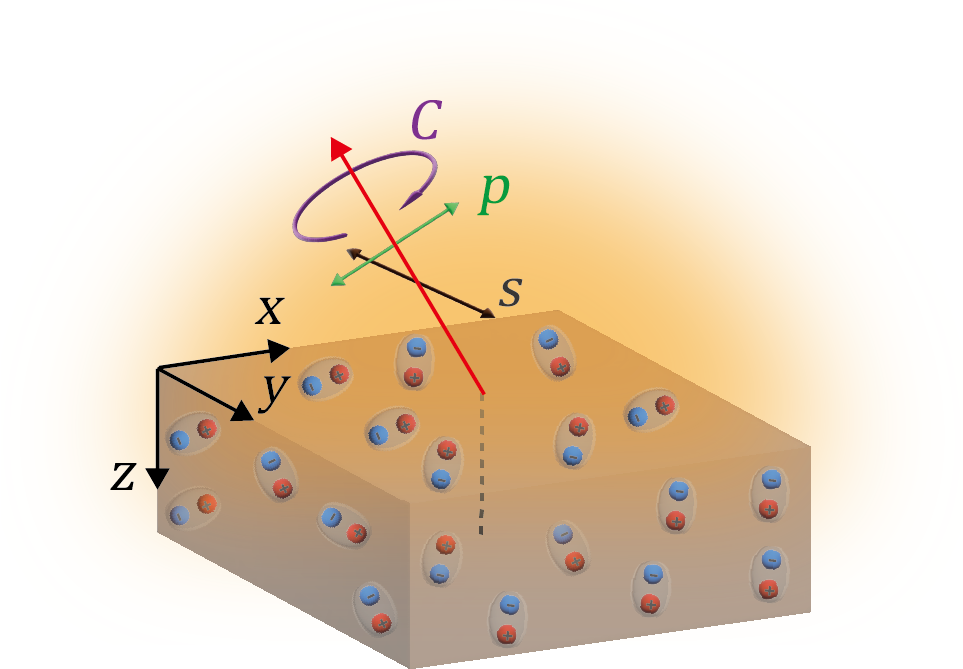}
\caption{Schematic of the semi-infinite anisotropic medium with partially polarized thermal emission. Thermal emission could be $p$-, $s$-, or circularly polarized depending on the direction of emission. With the assistance of DGFs, fluctuational electrodynamics can fully characterize the thermal emission generated by the media, providing a direct method to calculate the polarization related angular and spectral emissivities.}
\label{fig.1}
\end{figure}

\subsection{Far-field Stokes parameters and polarized emissivity}
As shown in Fig. \ref{fig.1}, the far-field region is of interest where the emission is assumed as collimated propagating waves \cite{RN1}. The medium is assumed at a uniform temperature and fills the $z > 0$ region with an interface at $z = 0$. The semi-infinite medium does not need to be homogeneous and may be made of layered anisotropic materials. The correlation function given in Eq. (\ref{eq6}) is composed of all possible wavevectors, including both propagating and evanescent waves. To determine Stokes parameters and calculate the angular and polarization-dependent emissivity, the correlation function should be decomposed into the wavevector space that corresponds a propagating wave toward the direction of emission. Since variables are translation invariant in the $x$-$y$ plane, the local frequency-domain electric field may be written as an integration over the wavevector space \cite{RN2, RN28}:
%%%%%%%%%%%%%%%%%%%%%%%%%%%
\begin{equation} \label{eq7}
\begin{split}
{\bf{E}}({\bf{r}},\omega ) = \int {\frac{{d{{\bf{k}}_\parallel }}}{{{{(2\pi )}^2}}}{\bf{E}}(z,\omega ,{{\bf{k}}_\parallel }){{{e}}^{i{{\bf{k}}_\parallel } \cdot {\bf{R}}}}} 
\end{split}
\end{equation}	 
%%%%%%%%%%%%%%%%%%%%%%%%%%
where $\bf{R}$ is the axial variables, ${\bf{k}}_\parallel$ is the wave vector projection onto the $x$-$y$ plane given by ${\bf{k}} = {{\bf{k}}_\parallel } + {k_z}{\bf{\hat z}}$ , ${k_z} = \sqrt {{k^2} - k_\parallel ^2}$ is the $z$-component of the wave vector. For propagating waves, ${k_z}$ is a real number. The DGF can also be decomposed according to \cite{RN2, RN18}:
%%%%%%%%%%%%%%%%%%%%%%%%%%%
\begin{equation} \label{eq8}
\begin{split}
\tensor{G}({{\bf{r}}_1},{{\bf{r}}_2},\omega ) = \int {\frac{{d{{\bf{k}}_\parallel }}}{{{{(2\pi )}^2}}}\tensor{g} ({z_1},{z_2},{{\bf{k}}_\parallel },\omega ){{{e}}^{i{{\bf{k}}_\parallel } \cdot ({{\bf{R}}_1} - {{\bf{R}}_2})}}} 
\end{split}
\end{equation}	 
%%%%%%%%%%%%%%%%%%%%%%%%%%
Here, $\tensor{g}$ denotes the DGF in the wave vector space that can be obtained based on Fresnel's coefficients as briefly derived and expressed in Appendix \ref{appendix:A}. The coherency matrix of the emission generated by the medium in the vector space is written as
%%%%%%%%%%%%%%%%%%%%%%%%%%%
\begin{widetext}
\begin{equation} \label{eq9}
\begin{split}
\langle {{\bf{E}}(z,\omega ,{{\bf{k}}_\parallel }) \otimes {{\bf{E}}^*}(z,\omega ,{{{\bf{k}}}_\parallel' })} \rangle =&  \delta ({{\bf{k}}_\parallel } - {{\bf{k}}_\parallel' })  \left[ {\begin{array}{*{20}{c}}
{\left\langle {{E_x}E_x^*} \right\rangle }&{\left\langle {{E_x}E_y^*} \right\rangle }&{\left\langle {{E_x}E_z^*} \right\rangle }\\
{\left\langle {{E_y}E_x^*} \right\rangle }&{\left\langle {{E_y}E_y^*} \right\rangle }&{\left\langle {{E_y}E_z^*} \right\rangle }\\
{\left\langle {{E_z}E_x^*} \right\rangle }&{\left\langle {{E_z}E_y^*} \right\rangle }&{\left\langle {{E_z}E_z^*} \right\rangle }
\end{array}} \right] 
\end{split}
\end{equation}	 
\end{widetext}
%%%%%%%%%%%%%%%%%%%%%%%%%%
The components in Eq. (\ref{eq9}) are obtained by subtracting the vacuum contribution as derived in Appendix \ref{appendix:B} from the global equilibrium contribution as given in Appendix \ref{appendix:A}. The field coherency matrix is independent of $z$ because the term ${{{e}}^{{{i}}{k_z}z}}$ with a real $k_z$ will drop out when multiplied by its complex conjugate. In general, Eq. (\ref{eq9}) may have nine nonzero elements due to the arbitrary emission direction; however, the rank of this matrix is two since the electric field must lie in the plane normal to $\bf{k}$, i.e., the propagation direction. A coordinates transformation is made by setting $z'$-direction to the $\bf{k}$-direction according to the following, 
%%%%%%%%%%%%%%%%%%%%%%%%%%%
\begin{equation} \label{eq10}
\begin{split}
\left[ {\begin{array}{*{20}{c}}
{x'}\\
{y'}\\
{z'}
\end{array}} \right] = \left[ {\begin{array}{*{20}{c}}
{\cos \phi }&{\sin \phi }&0\\
{ - \sin \phi }&{\cos \phi }&0\\
0&0&0
\end{array}} \right]\left[ {\begin{array}{*{20}{c}}
{\cos \theta }&0&{ - \sin \theta }\\
0&1&0\\
{\sin \theta }&0&{\cos \theta }
\end{array}} \right]\left[ {\begin{array}{*{20}{c}}
x\\
y\\
z
\end{array}} \right]
\end{split}
\end{equation}	 
%%%%%%%%%%%%%%%%%%%%%%%%%%
where $\phi$ and $\theta$ are the azimuthal angle and the zenith angle in the spherical coordinates that correspond to the direction of $\bf{k}$, respectively. In the new coordinates, only the $x'$- and $y'$-components of the electric field are nonzero. Hence, the 3 $\times$ 3 coherency matrix in Eq. (\ref{eq9}) is therefore reduced to a 2 $\times$ 2 coherency matrix in the tilted coordinates.

The Stokes parameters and the Poincaré sphere are frequently used in the polarimetric analysis of electromagnetic waves. An important property of Stokes parameters is the additivity \cite{RN30}, whereby the Stokes parameters of two completely incoherent waves can be added to yield the Stokes parameters of the combined wave. An unpolarized wave can be interpreted as a combination of incoherent linearly polarized waves with polarization in all directions. As a result, Stokes parameters can be written in the form of the ensemble averages: 
%%%%%%%%%%%%%%%%%%%%%%%%%%%
\begin{equation} \label{eq11}
\begin{split}
\left[ {\begin{array}{*{20}{c}}
{{S_0}}\\
{{S_1}}\\
{{S_2}}\\
{{S_3}}
\end{array}} \right] = \left[ {\begin{array}{*{20}{c}}
{\left\langle {{E_{x'}}E_{x'}^*} \right\rangle  + \left\langle {{E_{y'}}E_{y'}^*} \right\rangle }\\
{\left\langle {{E_{x'}}E_{x'}^*} \right\rangle  - \left\langle {{E_{y'}}E_{y'}^*} \right\rangle }\\
{\left\langle {{E_{x'}}E_{y'}^*} \right\rangle  + \left\langle {{E_{y'}}E_{x'}^*} \right\rangle }\\
{{{i}}\left\langle {{E_{x'}}E_{y'}^*} \right\rangle  - {{i}}\left\langle {{E_{y'}}E_{x'}^*} \right\rangle }
\end{array}} \right]
\end{split}
\end{equation}	 
%%%%%%%%%%%%%%%%%%%%%%%%%%
Though the unit of correlation functions in Eq. (\ref{eq11}) is different from that of the electric field squared (since the equation is written in ${\bf{k}}_\parallel$ space), it is termed as optical intensity hereafter for convenience. The Stokes parameters can be written in terms of the emissivity of a given polarization. The polarized emissivities are defined by the ratio of the optical intensity corresponding to the emission from the material to that from a blackbody as follows:
%%%%%%%%%%%%%%%%%%%%%%%%%%%
\begin{equation} \label{eq12}
\begin{split}
{\epsilon_j}(\omega ,{{\bf{k}}_\parallel }) = \frac{{\left\langle {{E_j}E_j^*} \right\rangle }}{{{S_{0,{\rm{bb}}}}(\omega ,{{\bf{k}}_\parallel })/2}}
\end{split}
\end{equation}	 
%%%%%%%%%%%%%%%%%%%%%%%%%%
Here, $j$ denotes the polarization of the electric field including $s$-polarized ${\bf{E}}_s = {\bf{E}}_{y'}$, $p$-polarized ${\bf{E}}_p = {\bf{E}}_{x'}$, left-hand polarized ${\bf{E}}_{\rm{L}} = ({\bf{E}}_{x'} + i{\bf{E}}_{y'})/\sqrt{2}$, right-hand polarized ${\bf{E}}_{\rm{R}} = ({\bf{E}}_{x'} - i{\bf{E}}_{y'})/\sqrt{2}$, ${{45}^ \circ }$ polarized ${\bf{E}}_{{45}^ \circ } = ({\bf{E}}_{x'} + {\bf{E}}_{y'})/\sqrt{2}$, and ${{135}^ \circ }$ polarized ${\bf{E}}_{{135}^ \circ } = ({\bf{E}}_{x'} - {\bf{E}}_{y'})/\sqrt{2}$.  ${S_{0,{\rm{bb}}}}(\omega ,{{\bf{k}}_\parallel })$ is the first Stokes parameter of the blackbody emission. Since thermal emission of the blackbody is unpolarized, a factor of $\frac{1}{2}$ is included in the optical intensity of blackbody to represent any particular polarization. The Stokes parameter or optical intensity of the blackbody is derived in  Appendix \ref{appendix:C} and the result is expressed as \cite{RN18, RN28}
%%%%%%%%%%%%%%%%%%%%%%%%%%%
\begin{equation} \label{eq13}
\begin{split}
{S_{0,{\rm{bb}}}}(\omega ,{{\bf{k}}_\parallel }) = {\left[ {\left\langle {{E_x}E_x^*} \right\rangle  + \left\langle {{E_y}E_y^*} \right\rangle } \right]_{{\rm{bb}}}} = 8\frac{{\pi {\mu _0}\omega }}{{{k_z}}}\Theta (\omega ,T)
\end{split}
\end{equation}	 
%%%%%%%%%%%%%%%%%%%%%%%%%%
Combining Eqs. (\ref{eq11})$-$(\ref{eq13}), the Stokes parameters can be expressed in terms of the polarized emissivities as \cite{RN31}
%%%%%%%%%%%%%%%%%%%%%%%%%%%
\begin{equation} \label{eq14}
\begin{split}
\left[ {\begin{array}{*{20}{c}}
{{S_0}}\\
{{S_1}}\\
{{S_2}}\\
{{S_3}}
\end{array}} \right] = \frac{{{S_{0,{\rm{bb}}}}(\omega ,{{\bf{k}}_\parallel })}}{2}\left[ {\begin{array}{*{20}{c}}
{{\epsilon_p} + {\epsilon_s}}\\
{{\epsilon_p} - {\epsilon_s}}\\
{{\epsilon_{{{45}^ \circ }}} - {\epsilon_{{{135}^ \circ }}}}\\
{{\epsilon_{\rm{R}}} - {\epsilon_{\rm{L}}}}
\end{array}} \right]
\end{split}
\end{equation}	 
%%%%%%%%%%%%%%%%%%%%%%%%%%
Once the Stokes parameters are obtained, the polarized emissivities can be solved using the identity $2{\epsilon_{\rm{avg}}}={\epsilon_{p}+{\epsilon_{s}} } = {\epsilon_{{{45}^ \circ }}}+{\epsilon_{{{135}^ \circ }}} ={\epsilon_{\rm{R}}}+{\epsilon_{\rm{L}}}$, where ${\epsilon_{\rm{avg}}}$ is the average emissivity given by the mean value of two emissivities for any orthogonal polarizations.

\section{Results and discusssion}

\subsection{Reciprocal multilayer structures}
\begin{figure}[!t]
  \includegraphics[width=86mm]{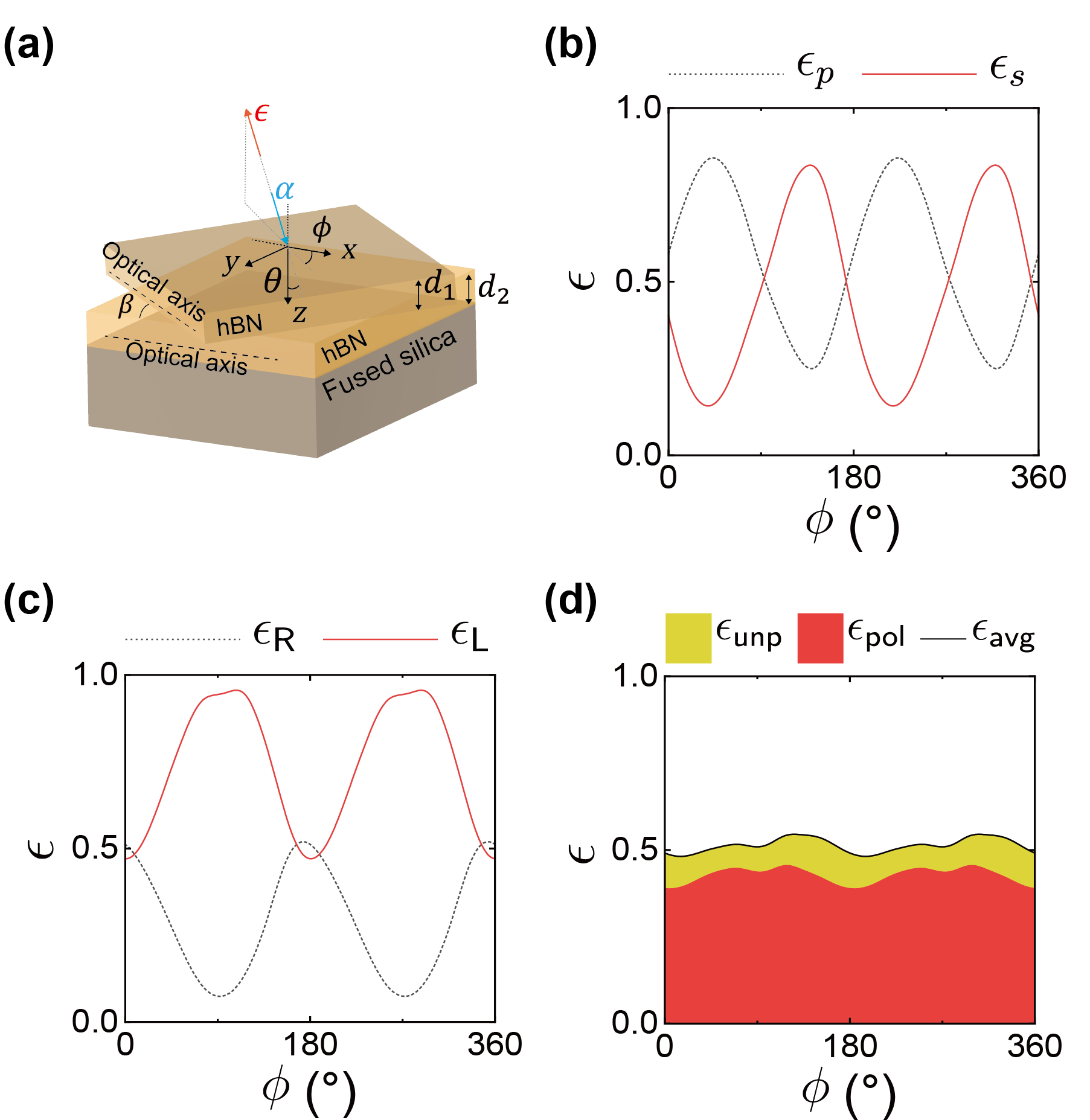}
  \caption{Reciprocal structure and the emissivities as a function of $\phi$ with $\theta = 52^\circ$ and frequency $\omega = 1587 $ $\rm{cm}^{-1}$. (a) Schematic of the reciprocal emitter based on the hBN-hBN-$\rm{SiO}_2$ structure: the optic axis of the top hBN layer ($d_1 = 0.62 $ ${ \mu}\rm{m}$) is in the $x$-$y$ plane at an angle $\beta = 45^\circ$ from the $x$-axis; the optic axis of the lower hBN layer ($d_2 = 2.66 $ ${ \mu}\rm{m}$) is parallel to the $x$-axis. The emissivities of (b) $p$ or $s$ polarizations, (c) L or R polarizations, and (d) average emissivities of polarized or unpolarized components.}
  \label{fig.2}
\end{figure}

It is well known that uniaxial materials can be used to achieve polarization conversion such as waveplates and selective transmission such as polarizers \cite{RN30}. Wu et al. \cite{RN32} showed that a bilayer structure made of two intertwisted hBN slabs could achieve tunable chirality with maximum circular dichroism as high as 0.84. The chiral response of the bilayer comes from the combined effects of selective transmission and polarization conversion. Since the circular dichroism of the proposed hBN bilayer was predicted at near normal incidence, it is expected that a large incidence angle may lead to a different optical response. The reciprocal emitter is designed based on the hBN bilayer structure as shown in Fig. \ref{fig.2}(a) with two hBN films on a fused silica ($\rm{SiO}_2$) substrate. Note that a single pair of variables $\theta$ and $\phi$ is used to denote directions of the absorptivity and the anti-paralleled emissivity, i.e., while the absorptivity $\alpha(\theta, \phi)$ is in the direction of $(\theta, \phi)$, emissivity  $\epsilon(\theta, \phi)$ points toward $(\pi-\theta, \phi+\pi)$. The optic axes of both the hBN layers are in the $x$-$y$ plane; however, the optic axis of the top layer is rotated about the $z$-axis by an angle of $\beta = 45^\circ$ with respect to the optic axis (parallel to the $x$-axis) of the lower hBN layer. The thicknesses of the hBN films are $d_1 = 0.62 $ ${ \mu}\rm{m}$ and $d_2 = 2.66 $ ${ \mu}\rm{m}$, respectively. Fused silica is chosen as the substrate due to its high emissivity at the frequency where hBN bilayer has a strong chiral response. The permittivity tensor of hBN is described by the Lorentz model with the parameters taken from Ref. \cite{RN33}, and the permittivity of the fused silica substrate is calculated using the optical constants from Ref. \cite{RN34}. 

For partially polarized emissions, Stokes parameters can be broken down into a completely polarized vector and an unpolarized vector, which are mutually independent \cite{RN30}:
%%%%%%%%%%%%%%%%%%%%%%%%%%%
\begin{equation} \label{eq15}
\begin{split}
{\bf{S}} = \left[ {\begin{array}{*{20}{c}}
{{S_0}}\\
{{S_1}}\\
{{S_2}}\\
{{S_3}}
\end{array}} \right] =  \left[ {\begin{array}{*{20}{c}}
{{S_{{\rm{pol}}}}}\\
{{S_1}}\\
{{S_2}}\\
{{S_3}}
\end{array}} \right] +\left[ {\begin{array}{*{20}{c}}
{{S_{{\rm{unp}}}}}\\
0\\
0\\
0
\end{array}} \right] 
\end{split}
\end{equation}	 
%%%%%%%%%%%%%%%%%%%%%%%%%%
where $S_{\rm{pol}} = {\sqrt {S_1^2 + S_2^2 + S_3^2} }$ and $S_{\rm{unp}} = {S_0 - S_{\rm{pol}}}$. The degree of polarization (DoP) is defined by
%%%%%%%%%%%%%%%%%%%%%%%%%%%
\begin{equation} \label{eq16}
\begin{split}
{\rm{DoP}} = \frac{{{S_{{\rm{pol}}}}}}{{{S_0}}} = \frac{{\sqrt {S_1^2 + S_2^2 + S_3^2} }}{{{S_0}}}
\end{split}
\end{equation}	 
%%%%%%%%%%%%%%%%%%%%%%%%%%
To investigate the angular dependence of the thermal emission from the structure, the emissivities are plotted as a function of $\phi$, at $\theta = 52^\circ$ and $\omega = 1587$ $ {\rm{cm}^{-1}}$, in Fig. \ref{fig.2}(b) for $p$ and $s$ linearly polarization and in Fig. \ref{fig.2}(c) for circularly polarization. In this work, the unit of $\omega$ is given as inverse centimeter for convenience. The actual emission angle is toward the negative $z$-direction. The frequency 1587 ${\rm{cm}^{-1}}$ corresponds to the edge of hyperbolic region with very low loss. The difference between the orthogonally polarized emissivities leads to the polarized portion of the thermal emission, such as the emissivity is largely $p$-polarized at $\phi = 45^\circ$ and $225^\circ$ (Fig. \ref{fig.2}(b)), and largely L-polarized at $\phi = 90^\circ$ and $270^\circ$ (Fig. \ref{fig.2}(c)). At around $\phi = 0^\circ$ and $180^\circ$, the two orthogonal emissivities in both Fig. \ref{fig.2}(b) and Fig. \ref{fig.2}(c) are almost identical. One may speculate that the emissivity is unpolarized. However, a sharp difference can be found between $\epsilon_{45^\circ}$ and $\epsilon_{135^\circ}$ at these azimuthal angles, which implies that the emissivity is largely $135^\circ$-polarized. In fact, emissivity can be largely polarized in any orientations depending on the azimuthal angle. Although an intuitive picture of thermal emission is provided by the polarized emissivities as shown in Fig. \ref{fig.2}(b) and Fig. \ref{fig.2}(c), it is inefficient to grasp the polarization information such as whether the emission is unpolarized or how to find the orientation with the largest polarized portion.

The average emissivities of polarized and unpolarized components are defined as
%%%%%%%%%%%%%%%%%%%%%%%%%%%
\begin{equation} \label{eq17}
\begin{split}
\begin{array}{l}
{\epsilon_{{\rm{pol}}}} = {\rm{DoP}} \cdot {\epsilon_{{\rm{avg}}}}\\
{\epsilon_{{\rm{unp}}}} = (1 - {\rm{DoP}}) \cdot {\epsilon_{{\rm{avg}}}}
\end{array}
\end{split}
\end{equation}	 
%%%%%%%%%%%%%%%%%%%%%%%%%%
Unlike the polarized emissivities, the average emissivities represent the magnitude of the polarized and unpolarized portions. Figure \ref{fig.2}(d) shows that the emission is largely polarized with $\epsilon_{\rm{pol}}$ around 0.4 and $\epsilon_{\rm{unp}}$ around 0.1 regardless of the azimuthal angle, though the polarization state is strongly dependent on the angle. In other words, one can always observe a significant difference between two orthogonal emissivities from this hBN bilayer structure.

\begin{figure}[!ht]
  \includegraphics[width=86mm]{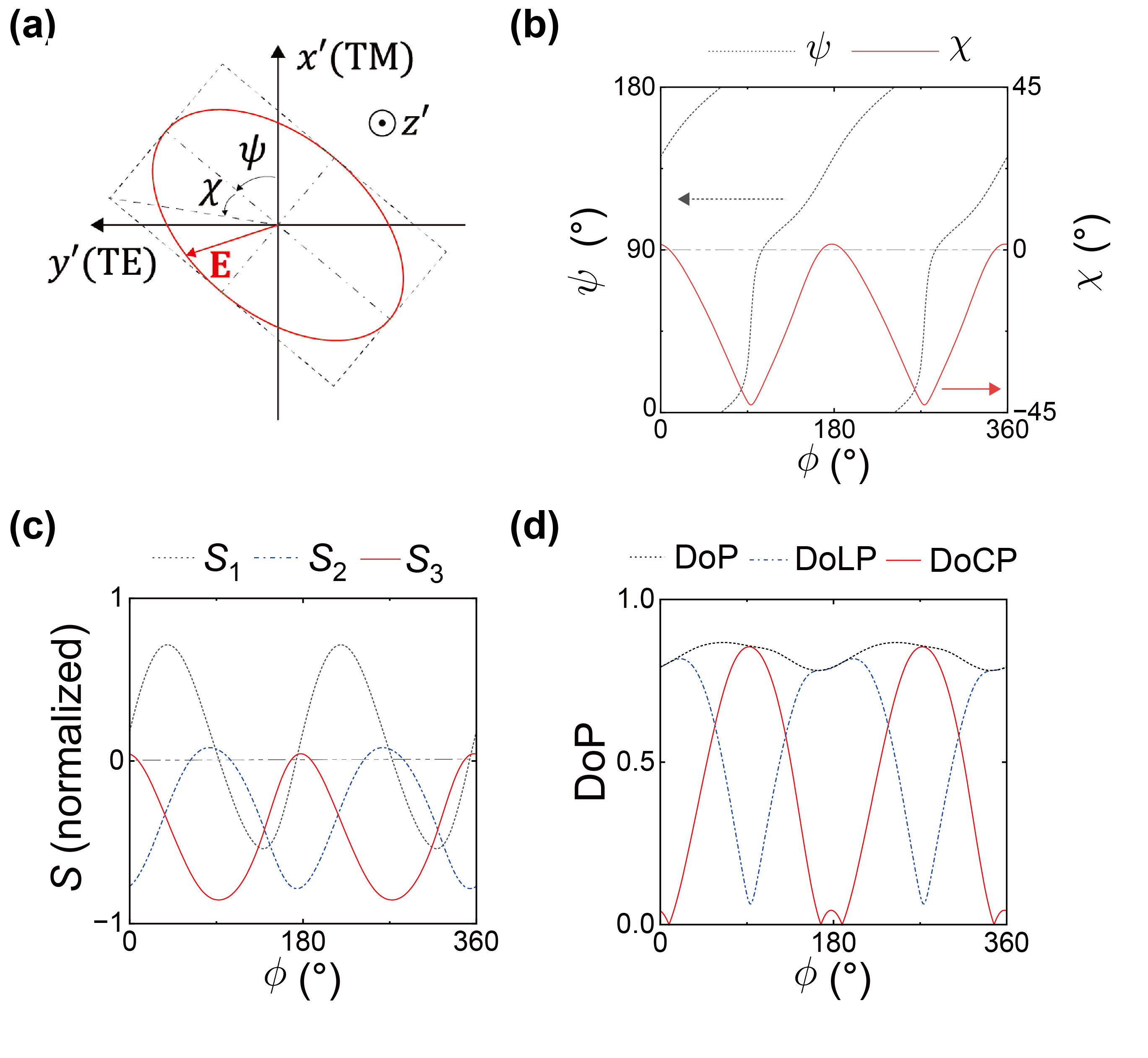}
  \caption{The polarization ellipse and the parameters related to the polarization state as a function of $\phi$ at $\theta = 52^\circ$ and $\omega = 1587 $ ${\rm{cm}^{-1}}$ based on the hBN-hBN-$\rm{SiO}_2$ structure. (a) Schematic of polarization ellipse. (b) Rotation angle $\psi$ $(0^\circ < \psi < 180^\circ)$ or the ellipticity angle $\chi$ $(-45^\circ < \chi < 45^\circ)$. (c) Polarization related Stokes parameters. (d) DoP, DoLP, or DoCP.}
  \label{fig.3}
\end{figure}

The unpolarized portion of the emission can be sufficiently depicted by a single parameter ${\epsilon_{\rm{unp}}}$. However, to fully specify the polarized portion, one needs to obtain the characteristic parameters of a polarization ellipse as shown in Fig. \ref{fig.3}(a). The distance between the center and the ellipse corresponds to the magnitude of the electric field as the wave travels. The polarization state is determined by two characteristic angles: the rotation angle $\psi$$(0^\circ < \psi < 180^\circ)$, which is the angle between $x'$-axis and the major axis of ellipse, and the ellipticity angle $\chi$$(-45^\circ < \chi < 45^\circ)$ \cite{RN30,RN35}. For example, a linearly polarized emission has $\chi=0^\circ$ with $\psi$ denoting the orientation of linear polarization; a circularly polarized emission has $\chi = \pm 45^\circ$ with an arbitrary $\psi$. The relationship between Stokes parameters and the characteristic angles is given by \cite{RN30}
\begin{figure}[!t]
  \includegraphics[width=86mm]{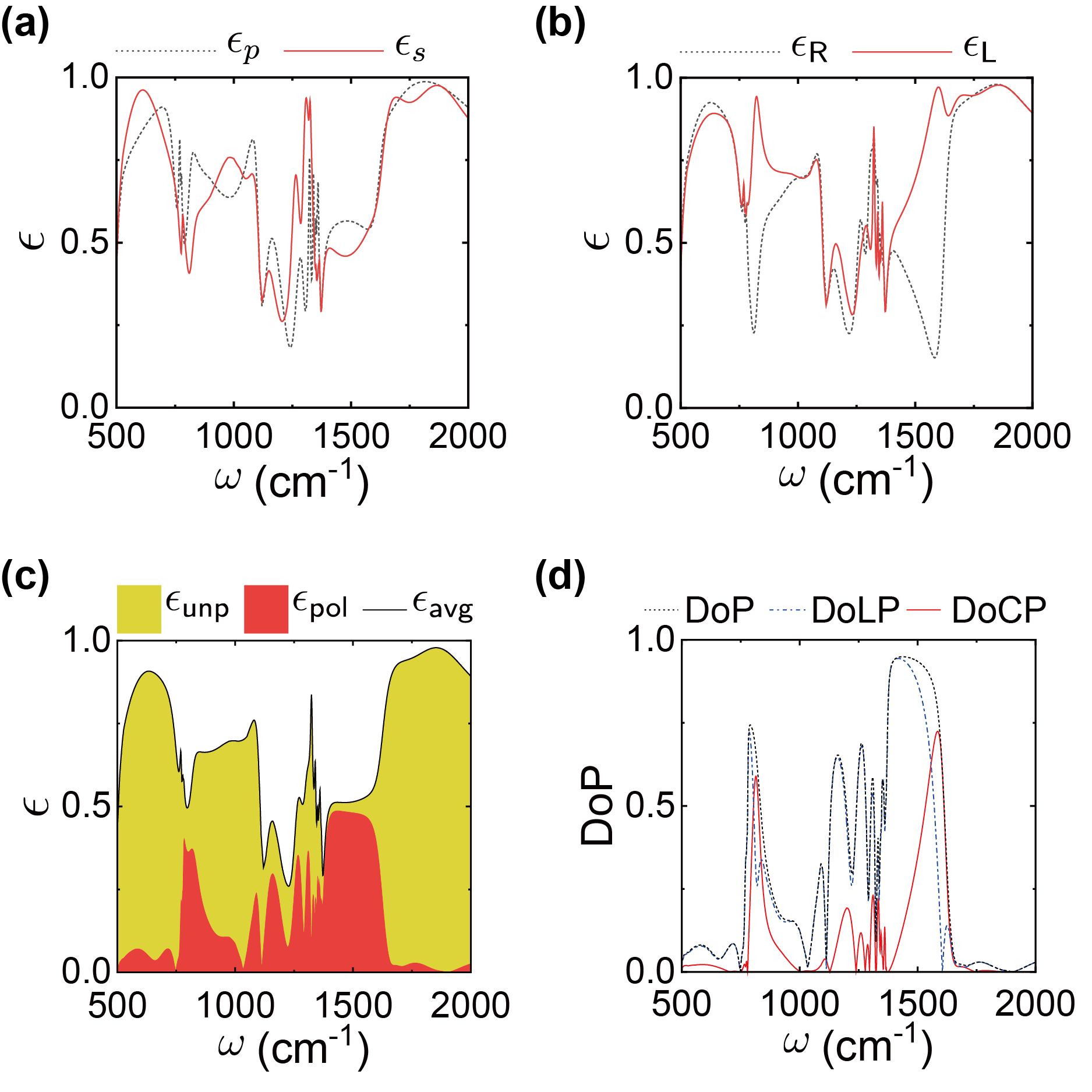}
  \caption{The polarized emissivities, average emissivities, and DoP, as a function of $\omega$ at $\theta = \phi = 0^\circ$ based on the hBN-hBN-$\rm{SiO}_2$ structure. Emissivities of (a) $p$ or $s$ polarizations, (b) L or R polarizations, (c) average emissivities of polarized or unpolarized components, and (d) DoP, DoLP, or DoCP.}
  \label{fig.4}
\end{figure}
%%%%%%%%%%%%%%%%%%%%%%%%%%%
\begin{equation} \label{eq18}
\begin{split}
\begin{array}{l}
\left[ {\begin{array}{*{20}{c}}
{{S_{{\rm{pol}}}}}\\
{{S_1}}\\
{{S_2}}\\
{{S_3}}
\end{array}} \right] = {\rm{DoP}} \cdot {S_0} \cdot \left[ {\begin{array}{*{20}{c}}
1\\
{\cos 2\psi \cos 2\chi }\\
{\sin 2\psi \cos 2\chi }\\
{\sin 2\chi }
\end{array}} \right]
\end{array}
\end{split}
\end{equation}	 
%%%%%%%%%%%%%%%%%%%%%%%%%%
Figure \ref{fig.3}(b) shows $\psi$ and $\chi$ as a function of $\phi$ for the same structure and conditions given in Fig. \ref{fig.2}. Note that $\psi=0^\circ$ and $180^\circ$ denote the same the orientations of major axis of the polarization ellipse. The combination of these angles reveals the polarization state of the polarized portion. Four independent parameters (average emissivity, DoP, and two characteristic angles in the polarization ellipse) provide all information to fully characterize the polarization state of thermal emission in a way similar to the four Stokes parameters. Figure \ref{fig.3}(c) plots the Stokes parameters for comparison. Here, the Stokes parameters are normalized to $S_0$. Though the calculations are based on a temperature $T = 300$ K, the results presented in this study are normalized and are not explicitly dependent on temperature. It should be noted that the optical properties of the materials are temperature dependent though only room temperature is considered in the present study. As shown in Fig. \ref{fig.3}(d), one can also use degree of linear polarization ${\rm{DoLP}} = \sqrt {S_1^2 + S_2^2} /{S_0}$ and degree of circular polarization ${\rm{DoCP}} =  |{S_3}| /{S_0}$ \cite{RN35}, respectively, to specify the polarization state. The sum of their squares is the square of DoP. Notice that the DoLP and DoCP do not represent the energy ratio of linearly or circularly polarized portion since the polarization ellipse cannot be further separated, except for $\chi=0^\circ$ (only linear polarization) and $\chi = \pm 45^\circ$ (only L or R circularly polarization). 

The emissivities with the polarizations of $p$ and $s$, R and L, the average emissivities, and DoP, are plotted as a function of $\omega$ from 500 to 2000 $\rm{cm}^{-1}$ at emission direction normal to the surface ($\theta = \phi = 0^\circ$) as shown in Figs. \ref{fig.4}(a)$-$\ref{fig.4}(d), respectively. The average emissivity has a value higher than 0.85 at frequencies around 625 $\rm{cm}^{-1}$ or 1850 $\rm{cm}^{-1}$. A high emissivity close to 1 suggests a nearly unpolarized emission with a small DoP. An average emission of unity corresponds to the blackbody limit since blackbody emission is incoherent and unpolarized. On the other hand, when the emission is completely polarized (DoP = 1), the average emissivity becomes 0.5, since the orthogonal component must be zero for a completely polarized emission. An example is when $\omega = 1433$ $\rm{cm}^{-1}$, the emitted waves in the normal direction is nearly linearly polarized with the DoP $\approx$ 0.95, $\epsilon_{\rm{avg}} \approx 0.51$, and the polarization orientation $\psi = 137^\circ$ given by the polarization ellipse. 

Since the material is reciprocal, the angular emissivity must equal the angular absorptivity in the antiparallel direction according to Kirchhoff's law. This is not always true for nonreciprocal systems, as to be discussed in the following. 
\begin{figure}[b!]
  \includegraphics[width=86mm]{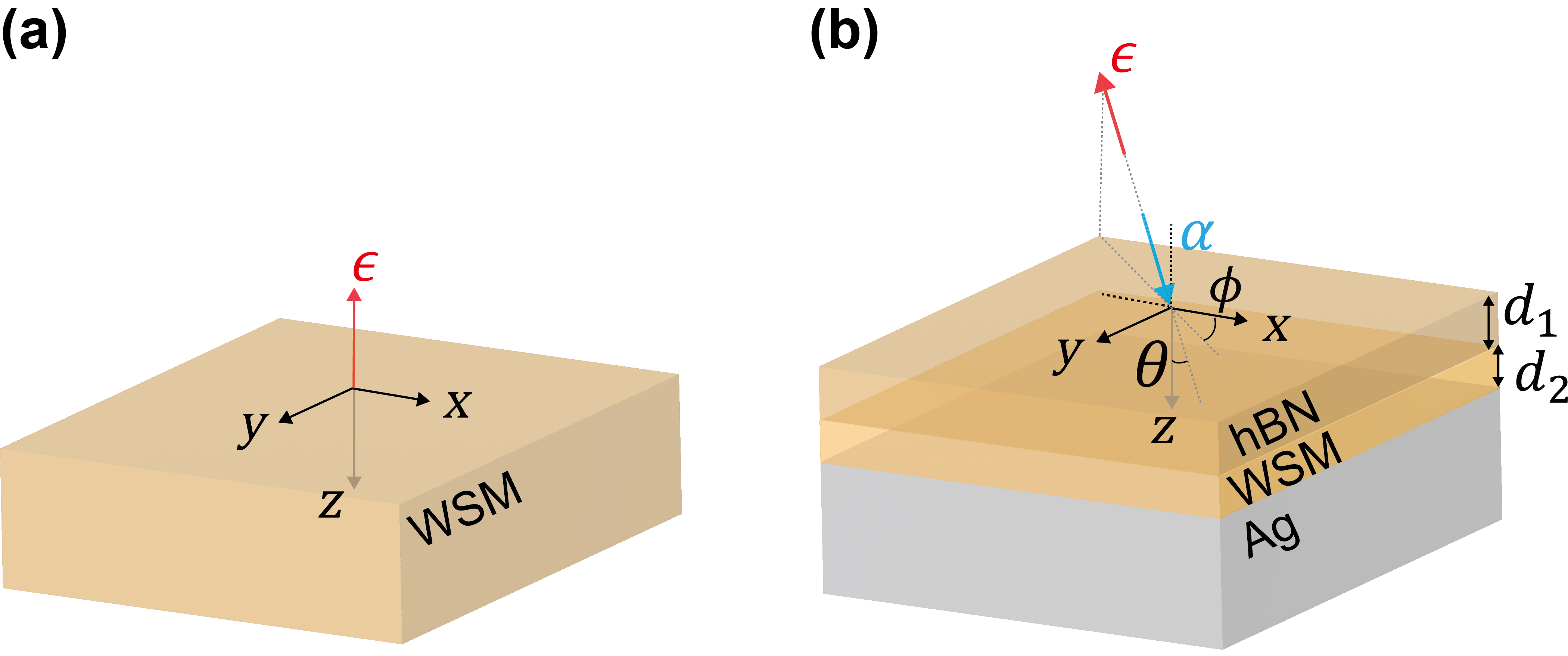}
  \caption{Schematic of WSM half-space and hBN-WSM-Ag nonreciprocal structure. (a) Semi-infinite WSM half-space with $\bf{b}$ along the $k_z$ direction. (b) The nonreciprocal hBN-WSM-Ag structure: the top hBN layer has $d_1 = 2.16$ ${\mu}\rm{m}$ with optic axis in ${\bf{\hat x}} + {\bf{\hat z}}$ direction; the middle WSM layers has $d_2 = 0.57$ ${\mu}\rm{m}$ with ${\bf{\hat b}} = \left( {{\bf{\hat x}} + {\bf{\hat y}} + {\bf{\hat z}}} \right)/\sqrt 3 $.}
  \label{fig.5}
\end{figure}

\subsection{Nonreciprocal structures}

Nonreciprocal emitters made of, for example, magneto-optical materials with asymmetric permittivity tensors break down the Lorentz reciprocity and consequently violate Kirchhoff's law \cite{RN11, RN12, RN13}. However, due to the weak magnetic response in the infrared range, the violation of Kirchhoff's law can be hardly observed with a bulk magneto-optical material. Grating structures that excite surface plasmon polaritons have been suggested to enhance the inequivalence between absorptivity and emissivity \cite{RN11, RN13, RN36}, and experimentally demonstrated using doped InAs gratings with an external magnetic field \cite{RN37}. Recent studies show that the magnetic Weyl semimetal (WSM) possesses significant potential in nonreciprocal emitter due to its giant magneto-optical effect \cite{RN13, RN38}. Consequently, the gyrotropic effect can be of two orders greater than that for traditional magneto-optical materials in a broad infrared region. It also grants WSM the ability to radiate strong circularly polarized thermal emissions \cite{RN39} or function as a circular polarizer \cite{RN40}. 

In energy harvesting applications, efficiency is limited by Kirchhoff's law since a high absorption indicates a high emission. Nonreciprocal emission/absorption is required to break the constrain and further increase the efficiency \cite{RN41}. Therefore, the design of nonreciprocal emitters should characterize the emissivity and absorptivity separately. Note that emissivity and absorptivity are not independent since specific equivalence can be found depending on the configurations of the emitters. For instance, Khandekar et al. \cite{RN15} discussed the modified Kirchhoff's law based on the circular polarization. Zhang et al. \cite{RN12} derived the modified Kirchhoff's law based on the co- and cross-polarization components of the reflectivity; a general relationship regardless of the reciprocity for a specular surface without transmission is given in the following: 
%%%%%%%%%%%%%%%%%%%%%%%%%%%
\begin{equation} \label{eq19}
\begin{split}
{\epsilon_{\rm{avg}}}(\omega ,\theta ,\phi ) = {\alpha _{{\rm{avg}}}}(\omega ,\theta ,\phi  + \pi )
\end{split}
\end{equation}	 
%%%%%%%%%%%%%%%%%%%%%%%%%%
which suggests that average emissivity and absorptivity in the direction symmetric to surface normal are always equal. In contrast, the relationship between angular emissivity and angular absorptivity for one polarization could be nonexistent. Guo et al. \cite{RN16} discussed these relationships with the help of the adjoint Kirchhoff's law, which states that the emissivity of an emitter equals the absorptivity of its mutually adjoint emitter obtained from a special transformation. 

Here, the analysis of nonreciprocal structures and justification of modified Kirchhoff's laws are carried out based on two nonreciprocal nanostructures as illustrated in Fig. \ref{fig.5}. Figure \ref{fig.5}(a) shows a schematic of a semi-infinite WSM half-space with its momentum separation $\bf{b}$ along the $k_z$ direction. The parameters used for WSM at room temperature are taken from Zhao et al. \cite{RN13}. Figure \ref{fig.5}(b) is the schematic of a nonreciprocal structure composed of hBN, WSM, and silver. The top hBN layer has $d_1 = 2.16$ ${\mu}\rm{m}$ with optic axis in ${\bf{\hat x}} + {\bf{\hat z}}$ direction, the middle WSM layer has $d_2 = 0.57 $ $ {\mu}\rm{m}$ with the unit vector of momentum separation ${\bf{\hat b}} = \left( {{\bf{\hat x}} + {\bf{\hat y}} + {\bf{\hat z}}} \right)/\sqrt{3}$, and the bottom layer is the opaque silver substrate whose dielectric function is modeled using the Drude model \cite{RN42}.

\begin{figure}[t!]
  \includegraphics[width=86mm]{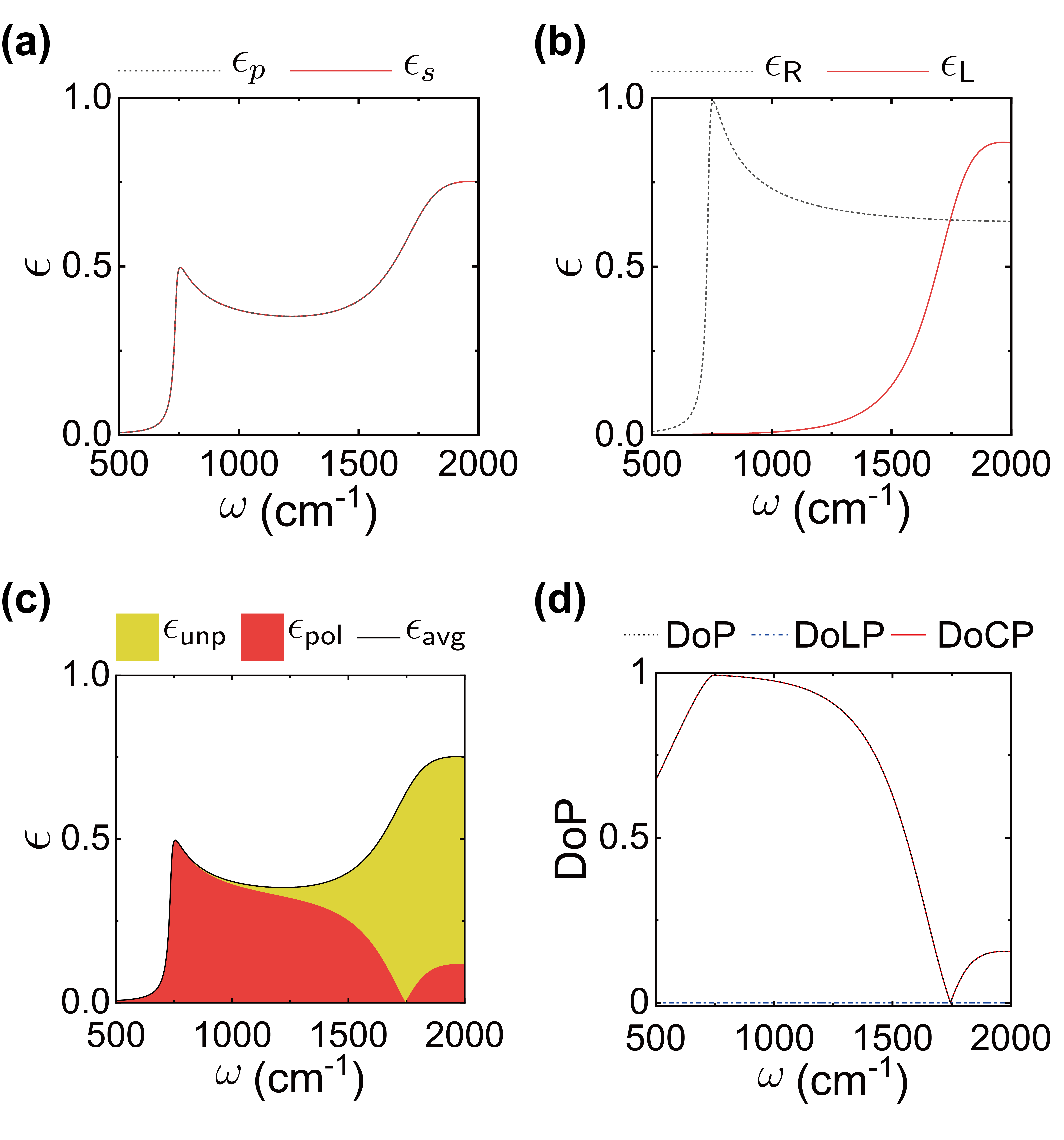}
  \caption{The emissivities and DoP as a function of $\omega$ at $\theta = \phi = 0^\circ$ for WSM half-space given in Fig. \ref{fig.5}(a). Emissivities of (a) $p$ or $s$ polarizations, (b) L or R polarizations, (c) average emissivities of polarized or unpolarized components, and (d) DoP, DoLP, or DoCP.}
  \label{fig.6}
\end{figure}

To begin with, the room-temperature emission of bulk WSM shown in Fig. \ref{fig.5}(a) in the normal direction is analyzed and the results are plotted in Fig. \ref{fig.6}. The linearly polarized emissivity of any orientation should be unchanged since $\bf{b}$ has no $x$- or $y$-component. Hence, the emissivity for $s$- or $p$-polarization is identify as shown in Fig. \ref{fig.6}(a). Figure \ref{fig.6}(b) plots circularly polarized emissivities, the distinction between the orthogonal emissivities indicates that the emission has a high DoCP. Figures \ref{fig.6}(c) and \ref{fig.6}(d) depict the average emissivities and the DoP, respectively. The average emissivity of the polarized portion maximizes at around $\omega = 757 $ $\rm{cm}^{-1}$, where $\epsilon_{\rm{avg}}\approx 0.50$ and DoCP $>$ 0.99. Unlike fused silica that has nearly unchanged optical constants in a wide temperature range, optical constants of WSM highly temperature-sensitive due to the change of the Fermi energy. Nevertheless, the effect of Weyl nodes separation and the number of Weyl nodes to the thermal emission of WSM, is analysis by Wang et al \cite{RN39}. According to the modified Kirchhoff's law for circular polarization, the bulk WSM has its circularly polarized emissivity equal to the absorptivity of orthogonal polarization (i.e., $\epsilon_{\rm{R}}=\alpha_{\rm{L}}$ and $\epsilon_{\rm{L}}=\alpha_{\rm{R}}$), which is also justified by the comparison with the absorptivity.

\begin{figure}[!ht]
  \includegraphics[width=75mm]{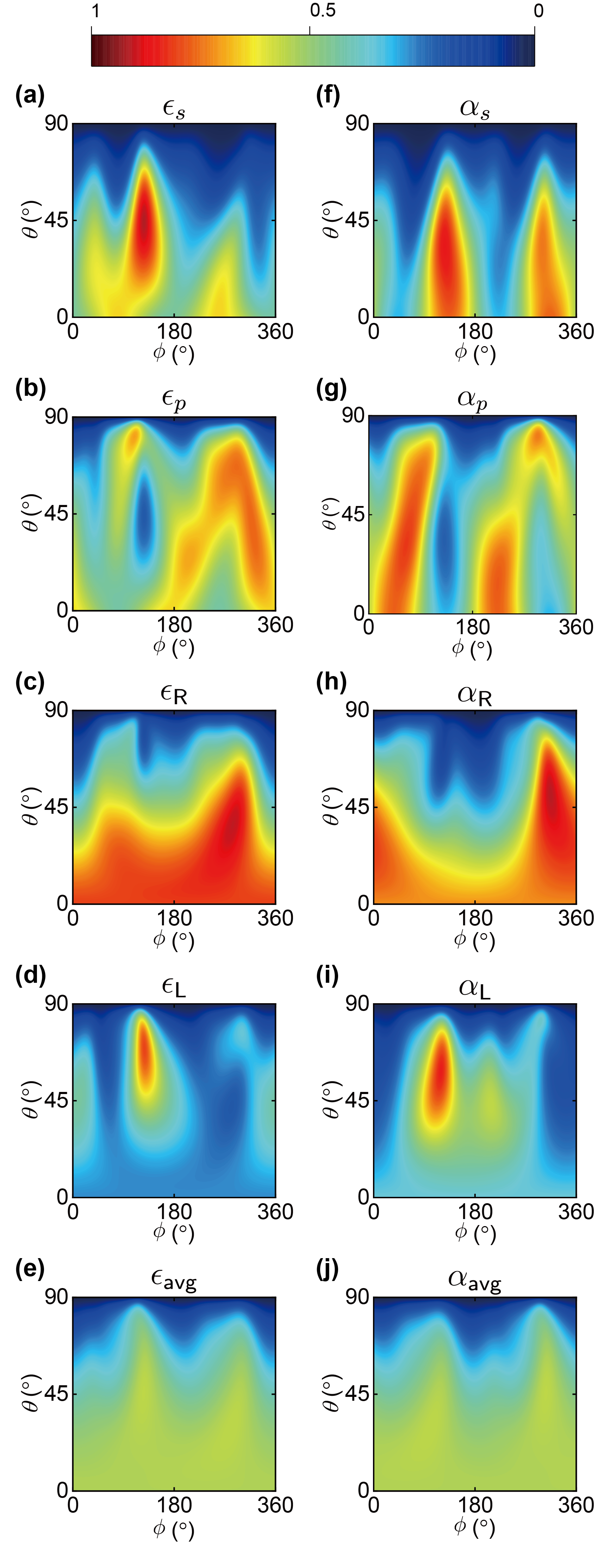}
  \caption{Contour plot of the emissivities and the absorptivities in the anti-parallel direction of the hBN-WSM-Ag structure at $\omega = 1316 $ $ \rm{cm}^{-1}$. (a)$-$(e) The emissivities of $s$-, $p$-, right-hand-, left-hand-polarizations, and average emissivity, respectively; (f)$-$(j) the polarized absorptivities of the same sequence.}
  \label{fig.7}
\end{figure}

The contour plot of the emissivities and the absorptivities in the antiparallel direction of the hBN-WSM-Ag structure shown in Fig. \ref{fig.5}(b) at $\omega = 1316 $ $\rm{cm}^{-1}$ are plotted in Fig. \ref{fig.7}. The emissivities of $s$-, $p$-, L-, R-polarizations, and the average emissivity, are shown in Figs. \ref{fig.7}(a)$-$\ref{fig.7}(e), respectively; the polarized absorptivities in the same sequence are shown in Figs. \ref{fig.7}(f)$-$\ref{fig.7}(j). Apparently, no relation exists between the polarized emissivity (Figs. \ref{fig.7}(a)$-$\ref{fig.7}(d)) and the absorptivity (Figs. \ref{fig.7}(f)$-$\ref{fig.7}(i)). For such kind of thermal emitters, it is impossible to obtain the emissivity by merely referring to its absorptivity. However, in terms of the average emissivity and absorptivity as shown in Figs. \ref{fig.7}(e) and \ref{fig.7}(j), one can relate them by using Eq. (\ref{eq19}) according to the modified Kirchhoff's law derived by Zhang et al \cite{RN12}. Furthermore, if the analysis is performed on the same geometry as Fig. \ref{fig.5}(b) but with a reversed $\bf{b}$ in the WSM layer (it is named as the adjoint emitter hereafter), one would observe that the initial polarized emissivity is equal to the adjoint polarized absorptivity in the antiparallel direction, and vice versa. Such observation is consistent with the adjoint Kirchhoff's law proposed by Guo et al \cite{RN16}. In short, the direct calculations using fluctuational electrodynamics lead to identical results as the modified Kirchhoff's law relations suggested in Refs. \cite{RN12, RN15, RN16}. Hence, the equivalence of the direct and indirect approaches in calculating the polarized emissivities are confirmed.

\begin{figure}[!ht]
  \includegraphics[width=86mm]{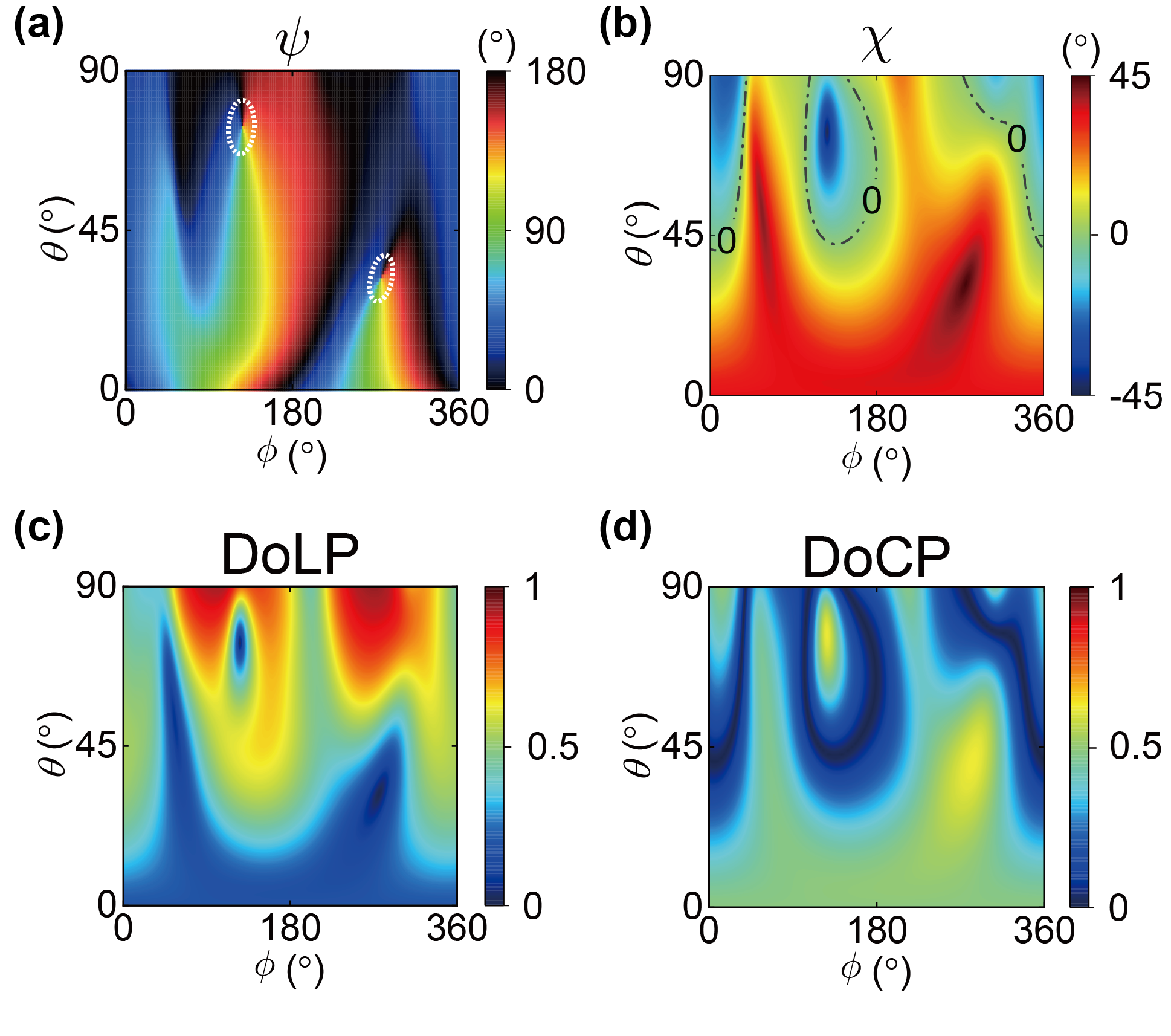}
  \caption{Parameters for the polarimetric analysis at $\omega = 1316 
$ $ \rm{cm}^{-1}$ based on the hBN-WSM-Ag structure. (a) Rotation angle $\psi$; (b) ellipticity angle $\chi$; (c) the DoLP and (d) the DoCP.}
  \label{fig.8}
\end{figure}

Emission characterization of the hBN-WSM-Ag structure needs a total of four independent Stokes parameters or their derived parameters such as $\psi$, $\chi$, DoP, and $\epsilon_{\rm{avg}}$. Contour plots of these parameters as functions of $\theta$ and $\phi$ are displayed in Fig. \ref{fig.8} to study the angular dependence. The rotation angle is plotted in Fig. \ref{fig.8}(a). Note that the color is set the same for $\psi = 0^\circ$ and $\psi =180^\circ$ due to their equivalent major axis orientation. Discontinuities in $\psi$ occur in the emission directions ($\theta, \phi$) near ($74^\circ, 125^\circ$) and ($32^\circ, 275^\circ$), as denoted by the dashed circles in Fig. \ref{fig.8}(a). This is because the lengths of minor and major axes of the polarization ellipse are equal, and a sudden change by $90^\circ$ may arise when the minor axis becomes the new major axis. In other words, the discontinuity appears when the polarized portion is circularly polarized with $\chi = \pm45^\circ$, as can be seen in the contour plot of $\chi$ in Fig. \ref{fig.8}(b). The contour line of $\chi = 0^\circ$ is labeled, indicates the polarized portion is completely linear. 

Polarimetric analysis can be used, for instance, to help design the structure with maximum polarized emission to fulfill the functionality of a nonreciprocal energy harvesting system. The analysis on $\psi$ and $\chi$ reveals the conditions for the desired polarization ellipse such as linear or circular polarization. However, it does not necessarily lead to a largely polarized emission due to the exclusion of the unpolarized portion. In contrast, the DoLP and DoCP are directly related to the linearly or circularly polarized emission, as shown in Fig. \ref{fig.8}(c) and Fig. \ref{fig.8}(d), respectively. For example, while $\chi = 45^\circ$ is achieved in the direction ($\theta, \phi$) = ($32^\circ, 275^\circ$), the DoCP is around 0.6, suggesting about 40\% of the emitted energy in this direction is unpolarized. In addition, the maximum DoCP is approximately 0.62 and occurs at the direction with $\theta \approx 42^\circ$ and $\phi \approx 283^\circ$, which does not coincide with the direction for $\chi = \pm45^\circ$. A maximum DoCP represents the largest ratio of $|S_3|/S_0$, which may generate emission with circular dichroism of the greatest ratio.

Generally speaking, the polarimetric analysis based on these parameters ($\psi$, $\chi$, DoCP, DoLP, and $\epsilon_{\rm{avg}}$) provides a further understanding of the polarization state. Though only a few selected examples are shown, it is expected that such analysis and understanding will help design polarization- and direction-selective nonreciprocal thermal emitters.

\section{Conclusion}
Fluctuational electrodynamics is applied as a direct approach to predict and analyze thermal emission from multilayered structures consisting of both reciprocal and nonreciprocal anisotropic materials. The direct calculation yields identical results as with the indirect method based on the modified Kirchhoff's law, therefore validating the appropriateness of both the direct and indirect methods. The Stokes parameters are introduced in the framework of fluctuational electrodynamics to carry out the polarimetric analysis. Characteristic parameters such as degree of polarization, rotation angle, ellipticity angle, etc., are used for describing the polarization state. Thermal emission from an intertwisted hBN bilayer on a fused silica substrate, a bulk WSM, a hBN/WSM/Ag structure are modeled. The results show that the emission could be circularly or linearly polarized in different emission directions. Furthermore, this work demonstrates that the preceding characteristic parameters can provide (i) sufficient information for polarization characterization and (ii) convenience in the designing process of the nonreciprocal emitters. This work will help the study of directional thermal emission and polarization control.

\begin{acknowledgments}
This work was supported by the National Science Foundation (CBET-2029892 for C.Y. and Z.M.Z.; DMR-2004749 for W.C.; and PHY-1748958 for Z.M.Z. to participate in a KITP summer program). Valuable discussions with Professor Shanhui Fan of Stanford University, Professor Mauro Antezza of the University of Montpellier, and Professors Marco Centini and Maria Cristina Larciprete of the University of Rome are greatly appreciated.
\end{acknowledgments}

\appendix

\section{Derivation of DGF}\label{appendix:A}
Assuming a point-like source in Eq. (\ref{eq1}) and substituting the current density with dipole moment, DGF can be written as 
%%%%%%%%%%%%%%%%%%%%%%%%%%%
\begin{equation} \label{eqA1}
\begin{split}
{\bf{E}}({{\bf{r}}_1},\omega ) = {\omega ^2}{\mu _0}\tensor{G} ({{\bf{r}}_1},{{\bf{r}}_2},\omega ) \cdot {\bf{p}}({{\bf{r}}_2},\omega )
\end{split}
\end{equation}	 
%%%%%%%%%%%%%%%%%%%%%%%%%%
where $\bf{p}$ is the dipole moment given by $\int {\bf{J}} {d}{\bf{r}} = {d}{\bf{p}}/{d}t$. Since the medium is semi-infinite with both ${\bf{r}}_1$ and ${\bf{r}}_2$ located in free space, the total field generated by the dipole moment can be decomposed into the direct portion, which arrives directly, and the reflected portion, which is reflected by the medium surface before arriving. The medium occupies the $z > 0$ half space, by letting $0 > z_1 > z_2$, the electric field can be written as an integration over the ${\bf{k}}_\parallel$ space as \cite{RN26}:
%%%%%%%%%%%%%%%%%%%%%%%%%%%
\begin{equation} \label{eqA2}
\begin{split}
{\bf{E}}({{\bf{r}}_1},\omega ) = & {\omega ^2}{\mu _0}\int {\frac{{{d}{{\bf{k}}_\parallel }}}{{{{(2\pi )}^2}}}} \frac{i}{{2{k_{0z}}}}{{{e}}^{i{{\bf{k}}_\parallel } \cdot ({{\bf{R}}_1} - {{\bf{R}}_2})}} \\
&\times\{ 
{{{e}}^{i{k_{0z}}({z_1} - {z_2})}}[ {{{{\bf{\hat e}}}_{s + }}({{{\bf{\hat e}}}_{s + }} \cdot {\bf{p}}) + {{{\bf{\hat e}}}_{p + }}({{{\bf{\hat e}}}_{p + }} \cdot {\bf{p}})} ] \\&
 + {{{e}}^{i{k_{0z}}( - {z_1} - {z_2})}}[ 
{r_{ss}}{{{\bf{\hat e}}}_{s - }}({{{\bf{\hat e}}}_{s + }} \cdot {\bf{p}})
 + {r_{sp}}{{{\bf{\hat e}}}_{p - }}({{{\bf{\hat e}}}_{s + }} \cdot {\bf{p}})\\&
 + {r_{pp}}{{{\bf{\hat e}}}_{p - }}({{{\bf{\hat e}}}_{p + }} \cdot {\bf{p}})
 + {r_{ps}}{{{\bf{\hat e}}}_{s - }}({{{\bf{\hat e}}}_{p + }} \cdot {\bf{p}}) ]
 \}
\end{split}
\end{equation}	 
%%%%%%%%%%%%%%%%%%%%%%%%%%
where $k_0$ is the wave vector in vacuum, $r_{jk}$ with $j, k = p, s$ is the Fresnel reflection coefficient; the first and second indices specify the polarization of the incident and reflected waves, respectively. Reflection coefficients are calculated using the modified 4$\times$4 transfer matrix method \cite{RN43}. The unit vectors of the polarized electric field are given by
%%%%%%%%%%%%%%%%%%%%%%%%%%%
\begin{equation} \label{eqA3}
\begin{split}
\begin{array}{l}
{{{\bf{\hat e}}}_{s \pm }} = {[ - \sin \phi ,\cos \phi ,0]^{\rm{T}}}\\
{{{\bf{\hat e}}}_{p \pm }} = 1/{k_0}{[ \pm {k_{z0}}\cos \phi , \pm {k_{z0}}\sin \phi , - {k_\parallel }]^{\rm{T}}}
\end{array}
\end{split}
\end{equation}	 
%%%%%%%%%%%%%%%%%%%%%%%%%%
Here, letting $z_1 < z_2$ will lead to a different electric field, but the result of coherency matrix will not be affected. Combining Eq. (\ref{eq8}), Eq. (\ref{eqA1}), and Eq. (\ref{eqA2}), one can obtain the following expression:
%%%%%%%%%%%%%%%%%%%%%%%%%%%
\begin{equation} \label{eqA4}
\begin{split}
\tensor{g} ({z_1},{z_2},{{\bf{k}}_\parallel },\omega ) = & \frac{i}{{2{k_{0z}}}}[ {{{e}}^{i{k_{0z}}({z_1} - {z_2})}}( 
{{{\bf{\hat e}}}_{s + }}{{{\bf{\hat e}}}_{s + }}^{\rm{T}}
 + {{{\bf{\hat e}}}_{p + }}{{{\bf{\hat e}}}_{p + }}^{\rm{T}}
 ) \\& + {{{e}}^{i{k_{0z}}( - {z_1} - {z_2})}}( 
{r_{ss}}{{{\bf{\hat e}}}_{s - }}{{{\bf{\hat e}}}_{s + }}^{\rm{T}}
 + {r_{sp}}{{{\bf{\hat e}}}_{p - }}{{{\bf{\hat e}}}_{s + }}^{\rm{T}}\\&
 + {r_{pp}}{{{\bf{\hat e}}}_{p - }}{{{\bf{\hat e}}}_{p + }}^{\rm{T}}
 + {r_{ps}}{{{\bf{\hat e}}}_{s - }}{{{\bf{\hat e}}}_{p + }}^{\rm{T}}
 ) ]
\end{split}
\end{equation}	 
%%%%%%%%%%%%%%%%%%%%%%%%%%
Follow the same procedure, $\tensor{g} ({z_2},{z_1},{{\bf{k}}_\parallel },\omega )$ can be determined by interchanging the source and the field while maintaining $z_1 > z_2$; therefore,
%%%%%%%%%%%%%%%%%%%%%%%%%%%
\begin{equation} \label{eqA5}
\begin{split}
\tensor{g} ({z_2},{z_1},{{\bf{k}}_\parallel },\omega ) =& \frac{i}{{2{k_{0z}}}}[ {{{e}}^{i{k_{0z}}({z_1} - {z_2})}}( 
{{{\bf{\hat e}}}_{s - }}{{{\bf{\hat e}}}_{s - }}^{\rm{T}}
 + {{{\bf{\hat e}}}_{p - }}{{{\bf{\hat e}}}_{p - }}^{\rm{T}}
 ) \\&+ {{{e}}^{i{k_{0z}}( - {z_2} - {z_1})}}( 
{r_{ss}}{{{\bf{\hat e}}}_{s - }}{{{\bf{\hat e}}}_{s + }}^{\rm{T}}
 + {r_{sp}}{{{\bf{\hat e}}}_{p - }}{{{\bf{\hat e}}}_{s + }}^{\rm{T}}\\&
 + {r_{pp}}{{{\bf{\hat e}}}_{p - }}{{{\bf{\hat e}}}_{p + }}^{\rm{T}}
 + {r_{ps}}{{{\bf{\hat e}}}_{s - }}{{{\bf{\hat e}}}_{p + }}^{\rm{T}}
 ) ]
\end{split}
\end{equation}	 
%%%%%%%%%%%%%%%%%%%%%%%%%%
The correlation function in the wave vector space at global thermal equilibrium becomes
%%%%%%%%%%%%%%%%%%%%%%%%%%%
\begin{widetext}
\begin{equation} \label{eqA6}
\begin{split}
\langle {{{\bf{E}}_{{\rm{ge}}}}({z_1},\omega ,{{\bf{k}}_\parallel }) \otimes {\bf{E}}_{{\rm{ge}}}^*({z_2},\omega ,{{\bf{k}}_\parallel' })} \rangle  = 16\pi \omega {\mu _0}\Theta (\omega ,T)\delta ({{\bf{k}}_\parallel } - {{{\bf{k}}}_\parallel' }) 
{\frac{{\tensor{g} ({z_1},{z_2},{{\bf{k}}_\parallel },\omega ) - {{\tensor{g} }^\dag }({z_2},{z_1},{{\bf{k}}_\parallel },\omega )}}{{2i}}}
\end{split}
\end{equation}	
\end{widetext}
%%%%%%%%%%%%%%%%%%%%%%%%%%
Note that a factor 4 is multiplied to the expression since only positive frequencies are considered.

\section{Derivation of the vacuum contribution}\label{appendix:B}
The vacuum part of the fluctuational electric field can be evaluated by considering an ideal blackbody surface located at $z = -\infty$. Only propagating waves arrive at the medium and then interact with the surface. The electric field of the vacuum part can be decomposed into the uncorrelated $p$ and $s$ components since the emission is unpolarized \cite{RN18, RN28}: 
%%%%%%%%%%%%%%%%%%%%%%%%%%%
\begin{equation} \label{eqB1}
\begin{split}
{{\bf{E}}_{{\rm{vac}}}}(\omega ,{\bf{r}}) = &\int \frac{{d{{\bf{k}}_\parallel }}}{{{{(2\pi )}^2}}}[{\bf{E}}_{{\rm{vac}}}^s(z,\omega ,{{\bf{k}}_\parallel }) + {\bf{E}}_{{\rm{vac}}}^p(z,\omega ,{{\bf{k}}_\parallel })] {{e}}^{i{{\bf{k}}_\parallel } \cdot {\bf{R}}}
\end{split}
\end{equation}	 
%%%%%%%%%%%%%%%%%%%%%%%%%%
where 
%%%%%%%%%%%%%%%%%%%%%%%%%%%
\begin{equation} \label{eqB2}
\begin{split}
{\bf{E}}_{{\rm{vac}}}^s(z,\omega ,{{\bf{k}}_\parallel }) = & {a_s}( {{{e}}^{i{k_{0z}}z}}{{{\bf{\hat e}}}_{s + }} + {r_{ss}}{{{e}}^{ - i{k_{0z}}z}}{{{\bf{\hat e}}}_{s - }}  \\&+ {r_{sp}}{{{e}}^{ - i{k_{0z}}z}}{{{\bf{\hat e}}}_{p - }} )\\
{\bf{E}}_{{\rm{vac}}}^p(z,\omega ,{{\bf{k}}_\parallel }) = & {a_p}( {{{e}}^{i{k_{0z}}z}}{{{\bf{\hat e}}}_{p + }} + {r_{ps}}{{{e}}^{ - i{k_{0z}}z}}{{{\bf{\hat e}}}_{s - }} \\&+ {r_{pp}}{{{e}}^{ - i{k_{0z}}z}}{{{\bf{\hat e}}}_{p - }} )
\end{split}
\end{equation}	 
%%%%%%%%%%%%%%%%%%%%%%%%%%
Here, $a_s$ and $a_p$ are the amplitudes for $s$- and $p$-polarized fields that satisfy the correlation relation:
%%%%%%%%%%%%%%%%%%%%%%%%%%%
\begin{equation} \label{eqB3}
\begin{split}
\left\langle {{a_j}a_k^*} \right\rangle  = \delta ({{\bf{k}}_\parallel } - {{\bf{k}}_\parallel' }){\delta _{jk}}\frac{{{S_{0,{\rm{bb}}}}(\omega ,{{\bf{k}}_\parallel })}}{2}
\end{split}
\end{equation}	 
%%%%%%%%%%%%%%%%%%%%%%%%%%
where ${S_{0,{\rm{bb}}}}(\omega ,{{\bf{k}}_\parallel })$ is given by Eq. (\ref{eq13}), $\delta_{jk}$ is the Kronecker delta that eliminates the cross-correlation terms. The above equations allow the evaluation of the vacuum contribution:
%%%%%%%%%%%%%%%%%%%%%%%%%%%
\begin{widetext}
\begin{equation} \label{eqB4}
\begin{split}
\langle {{\bf{E}}_{\rm{vac}}({z},\omega ,{{\bf{k}}_\parallel }) \otimes {{\bf{E}}_{\rm{vac}}^*}({z},\omega ,{{{\bf{k}}}_\parallel})} \rangle  = 
\langle {{{\bf{E}}_{\rm{vac}}^p}({z},\omega ,{{\bf{k}}_\parallel }) \otimes {{\bf{E}}_{\rm{vac}}^{p*}}({z},\omega ,{{{\bf{k}}}_\parallel })} \rangle 
 + \langle {{{\bf{E}}_{\rm{vac}}^{s}}(z,\omega ,{{\bf{k}}_\parallel }) \otimes {{\bf{E}}_{\rm{vac}}^{s*}}(z,\omega ,{{{\bf{k}}}_\parallel })} \rangle 
\end{split}
\end{equation}	 
\end{widetext}
%%%%%%%%%%%%%%%%%%%%%%%%%%

\section{Derivation of the blackbody optical intensity}\label{appendix:C}
One can evaluate the coherency matrix of a vacuum at global equilibrium by assuming DGFs have zero Fresnel reflection coefficients in Eq. (\ref{eqA6}). The result can be interpreted as two semi-infinite blackbody surfaces located at $z = \pm\infty$ that sandwich a vacuum in the middle. The coherency matrix from a single blackbody surface is half the magnitude at global thermal equilibrium. The Stokes parameter $S_{0,{\rm{bb}}}$ of the blackbody can be evaluated as: 
%%%%%%%%%%%%%%%%%%%%%%%%%%%
\begin{equation} \label{eqC1}
\begin{split}
{\delta ({{\bf{k}}_\parallel } - {{\bf{k}}_\parallel' })}
{S_{0,{\rm{bb}}}}(\omega ,{{\bf{k}}_\parallel }) = & \frac{1}{2}{\rm{Tr}} [ \langle {{\bf{E}}_{{\rm{ge}}}}({z_1},\omega ,{{\bf{k}}_\parallel }) \\ & \otimes {\bf{E}}_{{\rm{ge}}}^*({z_2},\omega ,{{{\bf{k}}}_\parallel' }) \rangle ]
\end{split}
\end{equation}	 
%%%%%%%%%%%%%%%%%%%%%%%%%%
Equation (\ref{eq13}) is obtained after simplification with zero reflection coefficients.

%\nocite{*}

\bibliography{prb}% Produces the bibliography via BibTeX.

\end{document}